==================

\documentstyle[preprint,revtex]{aps}

\def\prl#1{{\it Phys. Rev. Lett.} {\bf #1}}
\def\prr#1{{\it Phys. Rev.} {\bf #1}}
\def\prd#1{{\it Phys. Rev.} {\bf D#1}}
\begin{document}
\draft
\preprint{UBCTP93-010; July 1993}

\begin{title}
SCATTERING FROM A TWO DIMENSIONAL \\ARRAY OF FLUX TUBES: \\
A STUDY OF THE
VALIDITY OF MEAN FIELD THEORY
\end{title}

\vskip .5in

\author{Ken Kiers and Nathan Weiss}

\vskip 15pt

\begin{instit}
Department of Physics, University of British Columbia,\\
Vancouver, B.C. V6T 2A6, Canada
\end{instit}

\vskip .5in

\begin{abstract}
Mean Field Theory has been  extensively used in the study of systems of anyons
in two spatial dimensions. In this paper we study
the physical grounds for the validity of this approximation by considering
the Quantum Mechanical scattering of a charged particle from a
two dimensional array of magnetic flux tubes.
The flux tubes are
arranged on a regular lattice which is infinitely long in the ``$y$'' direction
but which has a (small) finite number of columns in the ``$x$'' direction.
Their physical size is assumed to be infinitesimally small.
We develop a method for computing the scattering angle as well as
the reflection and transmission coefficients to lowest order in the
Aharonov--Bohm interaction. The results of our calculation are compared
to the scattering
of the same particle from a region of constant magnetic field whose magnitude
is equal to the mean field of all the flux tubes.  For an incident plane
wave, the Mean Field approximation
is shown to be valid provided the flux in each tube is much less
than a single flux quantum.
This is precisely the regime in which Mean Field Theory for anyons
is expected to be valid.
When the flux per tube becomes of order 1, Mean Field Theory
is no longer valid.
\end{abstract}

\pacs{ }

\narrowtext
\section{Introduction}
\label{sec:intro}

There has been much interest in recent years in the phenomenon of
fractional statistics in two spatial dimensions.  Particles possessing
fractional statistics, known as anyons, besides being interesting in their
own right, have found applications in the Quantum Hall Effect \cite{qhe}
and have also been investigated in the context of
High $T_c$ Superconductivity.

Anyons are often studied in the framework of  ``Chern--Simons'' theory
in which they are modeled by ``attaching'' an infinitesimal
tube of ``statistical'' flux to each particle in the system.  In order to
calculate various properties of the anyon gas it is often assumed that
each anyon ``feels'' as if it were traveling in the mean magnetic
field due to all of the other anyons in the system. This ``Mean Field
Theory'' has been quite successful at describing many properties
of anyon systems including Anyon Superconductivity \cite{anyonsuper}
and the Fractional Quantum Hall Effect \cite{anyonfqhe}. In fact in
Chern--Simons Field Theory, the mean field approximation can be shown to be
valid when the statistical flux per particle is a small fraction of
a flux quantum.
Classically this Mean Field (MF) idea seems absurd
since the magnetic field is zero except at
isolated singularities and so there is no Lorentz force.  Quantum
mechanically however the vector potentials themselves attain
importance \cite{ab1}.  Thus it seems at least plausible that a
particle moving in a region populated by flux tubes would
in some respects behave as though it were in a mean
magnetic field.
 It is this idea which we investigate in this
paper.

It is quite straightforward to see that Mean Field Theory is at
best valid only when the flux per particle is not too large.
Assume
that a given anyon ``feels'' a mean field and
hence travels in a circular (Landau)  orbit.  If the area enclosed by the
orbit contains, on average, many other anyons then the approximation
may be valid.  Proceeding in this way one finds both for models based on
bosons \cite{bosonan} and on fermions \cite{ferman}
that the MF approximation is self-consistent only when the flux per
particle is much less than a single flux quantum.

It is our goal in this paper to examine the physical ideas implicit in the
above argument in more detail.  One question of interest in this
context is:
To what extent does a particle traveling in the presence of a number
of infinitesimally small flux tubes {\em really} behave as though it were in
some mean magnetic field?  In other words, to what extent does the
presence of the gauge field mimic the effect of a spatially
constant statistical
magnetic field?  Ideally we could consider an array of infinitesimally
small flux tubes and study the quantum mechanical motion of either
a charged particle or another flux tube in the presence of this array.
This however  turns out to be a very difficult problem to solve\cite{johnson}.
We choose instead to consider a specific configuration of flux tubes
consisting of a lattice which is of infinite
extent in the $y$ direction but which contains a (small) finite number of
columns in the $x$ direction.
A related problem (in which the array is random) has been considered previously
in
Ref. \cite{cm1} in which  scattering  from the array was considered
as an $incoherent$ sum of the scattering off each flux tube.
In this paper we consider the {\it coherent} scattering of a
charged particle incident on a lattice of flux tubes. The method which we
develop is
a variation on first order scattering theory.
Our main objective will be
to compute  the scattering angle and to compare
the result with  scattering from a strip of constant magnetic
field.

We thus begin in Section II by calculating the scattering angle of a plane
wave incident on an infinitely long strip of constant magnetic field.  In
Section III , after calculating some analytic results, we develop a formalism,
similar in spirit to the approach
conventionally used to
compute the index of refraction from the microscopic scattering amplitude
\cite{fermi,jackson},  to study  scattering of a charged particle from a
lattice
of flux tubes. We then use this formalism to find an expression for
 the scattering angle to first order in the interaction, which we then
compare to the analytic results.  (Note that
while exact solutions are available for  scattering from one
\cite{ab1} and, more recently, two \cite{guqian} flux tubes, the
problem of scattering from an infinite array of tubes cannot be completely
solved analytically.  We will find however that it is possible to solve
for the scattering angle exactly in a limited number of cases.)
In Sections II and III we also discuss the
reflection and transmission amplitudes.
Section IV contains a discussion of the results.

\section{SCATTERING FROM A STRIP OF CONSTANT MAGNETIC FIELD}
\label{sec:constfield}

In this section we briefly review the result
of scattering  a charged particle
from an infinitely long magnetic strip in two spatial dimensions.
We thus consider a situation in which the magnetic field $B$ is constant
in the region $0 \leq x \leq d$.  It is convenient to
work in a gauge in which $A_x = 0$ and
\equation
	A_y = \left\{ \begin{array}{l@{\hspace{.5in}}l}
			  -Bd/2,               & \mbox{if $x\leq 0$} \\~\\
			  Bx-Bd/2,          & \mbox{if $0\leq x\leq d$} \\~\\
			  Bd/2,	             & \mbox{if $x\geq d$}
			\end{array}
	      \right.
	\label{constfielda}
\endequation
The Schr\"{o}dinger equation for a particle with charge $e$ subject
to the above magnetic field is given by
\equation
	-{1\over 2m}[{\bf\nabla } -ie{\bf A}]^2 \psi (x,y) = E \psi (x,y).
	\label{schrod}
\endequation
This equation is solved by separating the $x$ and
$y$ dependence and representing the latter as a plane wave.  We thus
write
\equation
	\psi (x,y)=e^{i(k_y +{eBd\over 2})y} f(x)
\endequation
with
\equation
	{1\over 2m}[-{d^2 \over dx^2} +(k_y +{eBd \over 2}-A_y)^2] f(x)=
		E f(x).
\endequation
Suppose that the scattering particle is incident from the right (positive $x$).
The incident wave
\equation
	\psi _{\mbox{\scriptsize{inc}}} =e^{i(k_y + {eBd\over 2})y} e^{-ik_x x}
	\label{psiinc}
\endequation
will solve the Schr\"{o}dinger equation for $x>d$ provided
$2mE = k_{x}^2 + k_{y}^2$.

In the
interior region  ($0\leq x\leq d$) the particle behaves as
if it were in a potential which is a truncated parabola. Although
an explicit solution is in general not available, it is rather
straightforward to obtain a series solution.  For
certain combinations of the parameters
 the series truncates and the wave function is simply
a Hermite polynomial but in general the series is infinite.
(In the case when the series truncates, the wave function is
related to that of the Landau level.)
For $x$$<$$0$ the wave function is again given by a plane wave
\equation
	f(x)=e^{-i\tilde k_x x}
\endequation
but now
\equation
	2mE = \tilde k_{x}^2 + \left(k_{y}+eBd\right)^2
\endequation
It thus follows that for $x$$<$$0$
\equation
	\tilde k_x = \sqrt{k_{x}^2 -e^2 B^2 d^2 -2k_y eBd} \label{ktilde}
\endequation

Transmission of the particle to the region $x$$<$$0$ can occur if
the energy is sufficiently large so that
 $k_x^2 \geq e^2B^2d^2+2k_yeBd$. In this case
the solution to Eq. (\ref{schrod}) can  be
written as
\equation
	\psi = e^{i\left(k_y+{eBd\over 2}\right)y} \times \left\{ \begin{array}
		{l@{\hspace{.5in}}l}
		\tau ~ e^{-i\tilde k_x x}             &x\leq 0 \\
		~\\
		e^{-{\rm X} ^2 / 2}~ \left(c_0 G_0 ({\rm X}) +c_1 G_1 ({\rm X})\right)
		 &0\leq x\leq d \\
		~\\
		(e^{-ik_x x} +\rho ~e^{ik_x x}) &x\geq d
						       \end{array}
					       \right.
	\label{psigen}
\endequation
where $\tilde k_x$ is given by Eq. (\ref{ktilde}) and
\equation
	{\rm X} ~= ~ \sqrt{ |eB|}~\left(x-({k_y \over eB} +d)\right)
\endequation
is proportional to the distance from the centre of the parabola.
The functions $G_0$ and $G_1$ are given by
\begin{eqnarray}
	G_0 ({\rm X}) & = & 1 +{2(-\lambda )\over 2!}{\rm X}^2 +{2^2 (2-\lambda)
	(-\lambda)\over 4!} {\rm X}^4 + \cdots, 	\label{g0}\\~\nonumber\\
	G_1 ({\rm X}) & = & {\rm X} + {2(1-\lambda)\over 3!}
	{\rm X}^3 +{2^2(3-\lambda)(1-\lambda)\over 5!}{\rm X}^5 + \cdots, \label{g1}
\end{eqnarray}
\equation
	{\rm with}~~~~~~\lambda = {1\over 2}\left({k_x^2+k_y^2 \over  |eB|} - 1\right)
\endequation
$c_0$, $c_1$, $\rho$, and $\tau$ are constants which are fixed by requiring
that
the wave function and its derivative be continuous across the two boundaries.
$\rho$ and $\tau$ are the reflection and transmission amplitudes respectively.
Note that the series (\ref{g0}) and (\ref{g1}) converge for all ${\rm X}$.
(To see this note that for the high order terms in the expansion the ratio
of successive terms behaves like the corresponding ratio in the
expansion of exp($2{\rm X}^2$) \cite{arfken}.)
If the energy is small so that $k_x^2 <e^2B^2d^2+2k_y eBd$
there is no transmission and one obtains instead a decaying exponential for
$x<0$  so
that the wave is totally reflected.

It is quite straightforward
to compute the $angle$ by which the  particle is scattered.
The simplest way to do this is to compute the gauge invariant
current density given by
\equation
	{\bf J} = {1\over m}~{\rm Im}\left[ \psi^* \left({\bf \nabla }-
	ie{\bf A}\right) \psi \right]
\endequation
The angle $\phi$ which the current makes with the $x$-axis is then given by
\equation
	\mbox{tan}\phi = -{J_y\over J_x }.
\endequation
Using Eqs. (\ref{psiinc}) and (\ref{psigen}) we can compute the angle of both
the incident and the transmitted wave. We find that
\begin{eqnarray}
	\mbox{tan}\left(\phi_{\mbox{\scriptsize{inc}}}\right) & = &
	{k_y\over k_x} \nonumber\\
	\mbox{tan}\left(\phi_{\mbox{\scriptsize{trans}}}\right) & = &
	{k_y+eBd \over \sqrt{k_x^2 -e^2B^2d^2-2k_yeBd}}
	\label{phidefl}
\end{eqnarray}
This agrees exactly with the result obtained for the scattering angle
 in a purely classical treatment.
In fact the result simplifies
significantly when $|eB|d <<k_x$ in which case
\equation
{\mbox{tan}}\left(\phi_{\mbox{\scriptsize{trans}}}-\phi_{\mbox{\scriptsize{inc}}}\right)
	\simeq {eBd \over k_x} \label{deltaphi}
	\label{mainmf}
\endequation

{}From Eq. (\ref{phidefl}) we can understand the physical origin of the
restriction which was imposed above for $k_x^2$.
When $k_x^2 =e^2B^2d^2 +2k_yeBd$ the angle of the transmitted wave
$\phi_{\mbox{\scriptsize{trans}}}
=\pm \pi/2$ so that for $k_x^2 < e^2B^2d^2 +2k_yeBd$ the particle ``turns
around'' and
is re-emitted on the right.

It is evident from the above discussion that
in order to find $\phi_{\mbox{\scriptsize{trans}}}$ given $k_x$ and  $k_y$ it
is $not$
necessary to know the detailed form of the wave function in the interaction
region.
In fact all we actually used to get the scattering angle
in the above analysis was our knowledge of the vector potential on both
sides of the barrier and the fact that $k_y$ was conserved.
We shall come back to this point later.

We now proceed to a computation of the reflection and transmission
amplitudes $\rho$ and $\tau$.  In general exact values can only
be obtained numerically since the expressions will necessarily
contain the (generally infinite)
series $G_0 ({\rm X})$ and $G_1 ({\rm X})$ given in Eqs. (\ref{g0}) and
(\ref{g1})
and their derivatives.  It is however possible to obtain
approximate analytic expressions for $\rho$ and $\tau$ when the  magnetic field
is small by first relating
$G_0 ({\rm X})$ and $G_1 ({\rm X})$ to the confluent hypergeometric
functions as follows:
\begin{eqnarray}
	G_0 ({\rm X}) & = & M(-{\lambda \over 2},{1\over 2},{\rm X}^2),\label{g0m} \\
	G_1 ({\rm X}) & = & {\rm X} M({1\over 2}-{\lambda \over 2}, {3\over 2}, {\rm
X}^2),
	   \label{g1m}
\end{eqnarray}
where
\begin{eqnarray}
	M(a,b,z) & = & 1 + {az \over b} + {(a)_2 z^2\over (b)_2 2!}
		+ \cdots + {(a)_n z^n\over (b)_n n!} +\cdots, \\
	(a)_n & = & a(a+1)(a+2)\cdots (a+n-1), \\
	(a)_0 & = & 1.
\end{eqnarray}
The confluent hypergeometric functions are in turn related to the Bessel
functions by
\equation
	M(a,b,z)=\Gamma (b) e^{{1\over 2}z}({1\over 2}bz-az)^{{1\over 2}
	   -{1\over 2}b} \sum_{n=0}^{\infty}R_n ({z\over 2})^{n\over 2}
	   (b-2a)^{-{n\over 2}} J_{b-1+n}((2zb-4za)^{1\over 2}),
	\label{mbessel}
\endequation
where $R_0$$=$$1$, $R_1$$=$$0$, $R_2$$=$$b/2$, and
\equation
	R_{n+1} = [(n+b-1)R_{n-1}+(2a-b)R_{n-2}]/(n+1),
\endequation
for $n$$>$$2$ \cite{abst}.

We shall compute $\rho$ and $\tau$ for small $eBd^2$
only in the case
$k_y$$=$$0$.  To do this we shall need the values of $G_0 ({\rm X})$,
$G_1({\rm X})$ and their
derivatives at both $x$$=$$0$ and $x$$=$$d$.  When $x$$=$$d$,  X$=$$0$,
so these values are easily found by direct substitution into (\ref{g0}) and
(\ref{g1}) and the derivatives of these sums.  When $x$$=$$0$, however,
$X$$=$$-\sqrt{|eB|}d$, so we use (\ref{g0m}), (\ref{g1m}), and (\ref{mbessel})
to obtain expansions for $G_0$, $G_1$ and their derivatives
in powers of the quantity $\sqrt{|eB|}d$.  The calculations are considerably
simplified by the fact that ``$b$'' in (\ref{mbessel}) is always a half integer
so that the various Bessel functions reduce to  simple trigonometric
expressions.  Furthermore the arguments of the Bessel functions are simply
equal to ``$kd$''.  After a rather lengthy calculation we find,
for $|eB|d^2$$<<$$1$ and $k$$>>$$|eB|d$
\begin{eqnarray}
	\rho & = & {e^2B^2d^4\over 4}\left( {{\mbox{sin}}^2kd\over (kd)^4}
		+i ({{\mbox{sin}}2kd\over 2(kd)^4 } -{1\over (kd)^3})\right)
		+ O(e^3B^3d^6)	\label{rho}\\
	\tau & = & 1 + {e^2B^2d^4\over 4}\left( {1\over (kd)^2}-{2\over 3}{i
		\over kd}\right) +O(e^3B^3d^6) \label{tau}
\end{eqnarray}
Note that the above result satisfies the conservation of flux
\equation
	|\rho |^2 +{ {\tilde k} \over k}|\tau |^2 = 1
\endequation
to the required order, O($e^2B^2d^4$).

\section{SCATTERING FROM AN INFINITE STRIP OF FLUX TUBES}
\label{sec:fluxtubes}
\subsection{Analytic Results}

In this section we consider the scattering of a charged particle from a regular
array of infinitesimally small flux tubes. Our goal is to establish conditions
under which this scattering is equivalent to the scattering from a uniform
magnetic field whose magnitude is the mean field of all the flux tubes in the
array.

To this end consider a situation in which we have an
array of infinitesimally small flux tubes in two spatial dimensions each with a
flux
$\Phi$.  We take the lattice to be infinite in extent in the $y$ direction with
the
flux tubes separated by a distance $\xi$, but to have
$N+1$ ``columns'' in the $x$ direction with flux tubes positioned
at $x=0,\xi,\ldots ,N\xi $. We shall often choose to have only
one column of flux tubes ($N$$=$$0$). (The fact that we have set the spatial
separation
of the flux tubes in the $x$ and in the $y$ directions to be equal is simply
a convenience. Our results are easily generalizable to the case of unequal
spacing.)  We choose the gauge potential of each individual flux tube in
a radial gauge relative to the location of the flux tube so that
\equation
	A_r =0
	\nonumber
\endequation
\equation
	 A_\theta = {\Phi \over{2\pi r}} =-{\zeta \over {er}}
\endequation
where $r$ and $\theta$ are the coordinates relative
to the position of the flux tube and $\zeta$ provides a convenient
parameterization
of the flux.

The
contributions from each of the flux tubes in the lattice can be summed up to
obtain a closed
analytic result for the full gauge potential.
To do this one begins with the result \cite{brom}
\equation
	\mbox{cot}\pi z={1\over\pi}\sum_{n=-\infty}^{\infty} {1\over n+z},
\endequation
for complex $z$.  From this one can derive the formulae
\equation
	\sum_{m=-\infty}^{\infty} {1\over (m\alpha-\beta)^2 +\gamma^2}=
	 {i\pi \over 2\alpha\gamma}[\mbox{cot}\pi({\beta + i\gamma \over\alpha})-
	\mbox{cot}\pi({\beta - i\gamma
         \over \alpha})]
\endequation
and
\equation
	\sum_{m=-\infty}^{\infty} {m\alpha-\beta\over (m\alpha-\beta)^2 +\gamma^2}
	 =-{\pi \over 2\alpha}[\mbox{cot}\pi({\beta + i\gamma \over\alpha})+
	\mbox{cot}\pi({\beta - i\gamma
         \over \alpha})]
\endequation
where $\alpha$, $\beta$ and $\gamma$ are real.
Applying these results to our problem gives
\begin{eqnarray}
	A_x & = & {\zeta\pi \over e\xi} \sum_{n=0}^{N}
	{\mbox{sin}({2\pi y\over \xi}) \over \mbox{cosh}
	({2\pi \over \xi}(x-n\xi))-{\mbox{cos}}({2\pi y\over \xi}) }
	 \label{afieldx}\\
	~\nonumber \\
	A_y & = & -{\zeta\pi \over e\xi} \sum_{n=0}^{N} {\mbox{sinh}({2\pi
	\over \xi}(x-n\xi ))\over \mbox{cosh}({2\pi \over \xi}(x-n\xi))
	 -\mbox{cos}({2\pi y\over \xi}) }
	\label{afieldy}
\end{eqnarray}
where the sum is over the $N+1$ columns of the flux tube array. In fact when
$N$$=$$0$ the sum collapses to a single term.

Asymptotically (for large positive and negative $x$) the gauge potential
becomes
\equation
	\begin{array}{l@{\hspace{.5in}}l}
	A_x  \rightarrow  0 ~~~~{\rm as}~~~~ x \rightarrow \pm \infty \label{axasym}
	\\ ~\nonumber \\
	A_y  \rightarrow  \mp {\zeta\pi\over e\xi }(N+1) ~~~~{\rm as}~~~~
	 x \rightarrow \pm \infty
	\label{ayasym}
	\end{array}
\endequation
This asymptotic value is actually reached quite quickly. In fact the first
order
corrections damp exponentially as exp(-${2\pi \vert x\vert/\xi }$).
Not surprisingly the asymptotic value of the field is precisely that of
the constant magnetic field $B$ given in the previous section (Eq.
(\ref{constfielda})) with
\equation
	B~=~{{-2\zeta\pi}\over{e\xi^2}}
	\label{bmf}
\endequation
and
\equation
	d=(N+1)\xi
	\label{dmf}
\endequation
so that $B$ can be interpreted as the mean field of  the flux tubes.

One of our goals in this section is to compute the angle by which a charged
particle incident
from $x=+\infty$ is scattered by the array of flux tubes and to compare the
result
with the scattering angle off a strip of constant magnetic field. We thus wish
to compute
the transmitted angle $\phi_{\mbox{\scriptsize{trans}}}$ given $k_x$ and $k_y$
which can
be defined asymptotically as in the previous section. In the case of a strip of
constant
magnetic field we found (in the previous section)
that in order to compute the scattering angle it was
possible to ignore the  behaviour  of the vector potential and the
wave function in the region $0$$\le$$x$$\le$$d$ and to simply use
the asymptotic forms of these quantities and the fact that  $k_y$ was
conserved.
It is tempting to suggest that the same
result should apply in the present situation, in which case the scattering
angle for the array of flux tubes would be the same as that for the strip of
constant field.
This  result is however only true under certain restricted conditions.
The technical problem with the argument is that in our
case, due to the lack of (continuous) translational invariance in the $y$
direction,
the Schr\"{o}dinger equation does $not$ separate into $x$ and $y$ pieces.
The value of $k_y$ is thus not conserved through the ``interaction''
region of the potential.  The  {\em discrete} translational invariance
which is present in our case does, however, lead to interesting
predictions.  This will  be discussed below.

The simplest way to see that the above argument would lead to false conclusions
is to consider the case when the flux in each flux tube is an integer. In this
case the
corrections to the ``Mean Field'' result are large since the beam is in fact
totally undeflected.  This follows
from the fact that
the interaction can be completely ``gauged away'' by a nonsingular (except
possibly at the centres of the flux tubes) single--valued
gauge transformation.  In fact when $\zeta$ is an
integer the wave function
\equation
	\psi=e^{i(-k_x x+k_yy)}\mbox{exp}\left[i\zeta\sum_{n=0}^{N}\mbox{tan}^{-1}
	\left({e^{{2\pi\over\xi}(x-n\xi)}-
	\mbox{cos}({2\pi y\over\xi}) \over \mbox{sin}({2\pi y\over \xi})}\right)
	 -{i\zeta \pi y\over \xi}\right]
	\label{psiint}
\endequation
is single--valued and solves Eq. (\ref{schrod}) with ${\bf A}$ given by
 Eqs. (\ref{afieldx}) and
(\ref{afieldy}).  The resulting current has tan$\phi =k_y/k_x$ everywhere
(except
possibly at the singularities).
In cases where $\zeta$ is not an integer the wave function
(\ref{psiint}) is multi--valued and is therefore not
an admissible solution to the Schr\"{o}dinger equation.
These results were, of course, already noted in the original paper of
Aharonov and Bohm\cite{ab1}.
It is thus clear that replacing the flux tubes by a mean field will in general
give
an $incorrect$ answer. We shall study below the circumstances under which the
mean field
result $is$ correct.

As one might expect from the fact that there is no scattering when $\zeta$ is
an integer, the current (and hence the deflection angle) is a periodic function
of
 $\zeta$. To see this, suppose $\psi_{\zeta}$ is an exact solution to the
Schr\"{o}dinger
equation with $0<\zeta<1$.  Let $p$ be an integer and perform
a gauge transformation with gauge function
\equation
	\chi={p\over  e}\sum_{n=0}^{N}\mbox{tan}^{-1}\left(
	{e^{{2\pi\over\xi}(x-n\xi)}-
	\mbox{cos}({2\pi y\over\xi}) \over \mbox{sin}({2\pi y\over \xi})}
	\right) -{p \pi y\over e\xi},
\endequation
This now gives an admissible solution for the case $\zeta \rightarrow p+\zeta$.
The current is of course unchanged by the gauge transformation. Thus the
scattering
angle is periodic in $\zeta$ with a unit period.

It turns out that for certain values of $k_x$, $k_y$, and $\zeta$
we can solve for the scattering angle exactly without
knowing the form of the wavefunction in the scattering region\cite{remark}.
The solution
depends on the periodicity of the lattice in the $y$ direction \cite{modular}.
Following the usual discussion of Bloch's theorem \cite{bloch}, one can define
a
unitary translation operator $T_n$ as follows:
\equation
	T_n f(x,y) = f(x,y + n\xi).
\endequation
Since our Hamiltonian (with the vector potential defined in (\ref{afieldx}) and
(\ref{afieldy}) ) is invariant under $y\rightarrow y+n\xi$, we have the
operator identity
\equation
	[T_n , H ] = 0.
\endequation
Furthermore, the $T_n$'s form a representation of the group of translations
by $\xi$ so that
\equation
	T_nT_m=T_mT_n=T_{n+m},
\endequation
which means that eigenstates of $H$ may be chosen to be simultaneous
eigenstates
of {\em all} of the $T_n$:
\begin{eqnarray}
	H\psi & = & E\psi ,\\
	T_n \psi & = & c_n \psi ,
\end{eqnarray}
with $c_n c_m$$=$$c_{n+m}$.  Thus $c_n$$=$exp$(2\pi in\delta)$ for some real
parameter
$\delta$, which will be related to $k_y$ in this case.

We now consider the form of the wavefunction as $x\rightarrow \infty $ (the
incoming
wave).  In this region we may take (see Eq. (\ref{psigen}) )
\equation
	\psi (x,y)= e^{i(k_y -\zeta\pi(N+1)/\xi)y}(e^{-ik_x x} +\rho e^{ik_xx}),
	\:\:\:\:\: x\rightarrow \infty,
\endequation
so that
\equation
	\delta ={k_y \xi \over 2\pi} -{\zeta (N+1) \over 2}.
\endequation
To the far left of the scattering region we again seek a plane wave solution,
so we
set
\equation
	\psi (x,y)=\tau  e^{i(k_y^{'} -\zeta\pi(N+1)/\xi )y} e^{-i{\tilde k_x} x},
	\:\:\:\:\: x\rightarrow -\infty.
\endequation
The fact that $c_n$$=$exp$(2\pi in\delta)$ then gives
\equation
	k_y^{'} = k_y +{2\pi m \over \xi}.
\endequation
Energy conservation then determines $\tilde k_x$:
\equation
	\tilde k_x^m = [ k_x^2 -({2\pi \over \xi})^2(\zeta (N+1) - m)^2 +{4\pi k_y
\over\xi}(\zeta (N+1) - m)]^{1\over 2},
\endequation
so that if $\tilde k_x^m$ is real the angle of the beam of transmitted
particles is determined by the equation
\equation
	{\mbox{tan}} \phi_{\mbox{\scriptsize{trans}}}^m  ={k_y -{2\pi \over\xi}(\zeta
(N+1) -m)\over
	[ k_x^2 -({2\pi \over \xi})^2(\zeta (N+1) - m)^2 +{4\pi k_y \over\xi}(\zeta
(N+1) - m)]^{1\over 2}}.
	\label{phiexact}
\endequation
Comparing the above equation with Eqs. (\ref{phidefl}), (\ref{bmf}) and
(\ref{dmf}),
we see that agreement with the MF result is achieved when $m$$=$$0$.

We now restrict ourselves to the case where $N$$=$$0$ and consider plotting
tan$\phi_{\mbox{\scriptsize{trans}}}$ versus $\zeta$.  The ``critical points''
(i.e., the points for which $\tilde k_x^m$$=$$0$ ) are given by
\equation
	\zeta^{\pm}_m =m + {k_y \xi \over 2\pi} \pm (k_x^2+k_y^2)^{1\over 2}{\xi \over
2\pi}.
	\label{zetapm}
\endequation
If $(k_x^2+k_y^2)^{1\over 2}\xi$$\leq$$\pi$, the scattering angle is
well defined in the intervals $(\zeta^{-}_m,\zeta^{+}_m)$ and is given by Eq.
(\ref{phiexact}).
In the intervals$(\zeta^{+}_m,\zeta^{-}_{m+1})$, all the $\tilde k_x^m$ are
imaginary and the
beam of particles is backscattered by the flux tubes.  If
$(k_x^2+k_y^2)^{1\over 2}\xi$$>$$\pi$, there are regions of the $\zeta$ axis
for which more than one scattering angle could be defined.  Had  the incident
beam  been
characterized by a wave packet in our calculations,
the scattered particles would have split (asymptotically)  into several wave
packets with distinct
directions in space, given by (\ref{phiexact}).  Since our incident beam is
characterized
by a plane wave the scattered wave can never be  separated into
its various pieces and the scattering angle is not well defined.  If we view
the situation from
the perspective of diffraction theory,  we see that the ``circular waves''
being emitted from
each of the scattering centers are interfering constructively in more than one
direction.

A similar result holds for any value of $N$.  There may be certain values of
the
parameters for which there is no value of $m$ with real
$\tilde k_x^m$.  In such cases the beam of particles is totally reflected
by the flux tubes.  If there is only one integer $m$ for which $\tilde k_x^m$
is real
the scattering angle is well defined and is given by (\ref{phiexact}).  If, in
addition,
$m$$=$$0$, the angle agrees exactly with the MF case.  Finally, it is possible
that there
exist more than one integer $m$ for which $\tilde k_x^m$ is real.
This situation  corresponds to the ``diffraction'' case which was considered
above.

Finally,  note that the invariance of $\phi_{\mbox{\scriptsize{trans}}}$ under
$\zeta \rightarrow
 \zeta + m^{'} $ (which is expected in light of the gauge invariance arguments
given above)
is already included in Eq. (\ref{phiexact}) by  simply taking  $m \rightarrow m
-m^{'} (N+1)$.

\subsection{Numerical Results}

The above analytic analysis is somewhat limited in that it only predicts the
scattering angles and in that it depends in an essential way on the periodicity
of the lattice.  We shall now develop a numerical technique which will allow us
to
examine not only the scattering angle but also the lowest order values of the
reflection and transmission amplitudes.
This technique can be generalized to a non-periodic array of flux tubes.

As discussed in the introduction, we expect the Mean Field result
to be valid when the flux per ``particle'' $\zeta$ is small. We thus
analyze our problem for small $\zeta$.
Consider a wave which is incident from $x=+\infty$. This incident wave will
induce a scattered wave from each of the flux tubes in the lattice.
To linear order in
$\zeta$ the total wave function can be viewed as the incident wave plus the
sum of the scattered waves from each individual flux tube.
This procedure is, however, significantly
more complicated than in ordinary scattering theory due to the presence of the
Aharonov--Bohm phases.

In order to proceed we shall begin by  reviewing
the results for scattering from a $single$
flux tube which we assume is positioned at the origin.
In this case, an exact scattering solution
is known  \cite{ab1}. It is given by
\equation
	\psi = \sum_{m=-\infty}^{\infty} e^{-{i\pi\over 2}|m+\zeta |}
	J_{|m+\zeta |}(kr) e^{im\theta },
	\label{abwfn1}
\endequation
where $J$ is a Bessel function and the angle
$\theta$ should be taken from $-\pi$ to $\pi$ (where $\theta$$=$$0$ corresponds
to the positive $x$ axis). To see that this is a scattering solution
one notes that as $x\rightarrow\infty$
\equation
\psi \simeq e^{-ikx-i\zeta \theta}
\endequation
so that the incident current density is in the
negative $x$ direction.  In fact for $|\zeta |<1$, Aharonov and Bohm
(with modifications due to Hagen \cite{hagen}) showed that (\ref{abwfn1})
reduces
asymptotically (as $r\rightarrow\infty$) to
\equation
	\psi \rightarrow e^{-ikx -i\zeta \theta} \mp i{e^{ikr}\over (2\pi ikr
         )^{1/2}}{\mbox{sin}}\zeta\pi {e^{\mp i\theta /2}\over
	{\mbox{cos}}(\theta /2)},
	\label{abasym}
\endequation
where the upper (lower) sign holds for $\zeta$ positive (negative).
The above asymptotic expression for $\psi$ requires one further crucial
assumption
which is that $(kr(1\pm {\mbox{cos}}\theta))^{1/2}$ are large.  This
assumption fails in the forward and backward directions so that
(\ref{abasym}) is inaccurate in these regions.  This is fortunate
since (\ref{abwfn1}) is convergent and single--valued for all $\theta$
as can be confirmed by a direct
evaluation of  (\ref{abwfn1}) near $\theta$$=$$\pm\pi$.
The asymptotic form (\ref{abasym}), however, is not single valued and it
diverges when $\theta$$=$$\pm\pi$.

This ``splitting'' of the wave, in the case of a single scatterer, into
an incident and a scattered wave is essential for us since it is
the starting point for summing up the scattered waves from a $lattice$
of flux tubes. We cannot simply use Eq. (\ref{abasym}) for this
splitting, even though
the ``scattered'' wave is proportional to $\mbox{sin}(\zeta\pi)$ which is small
for small $\zeta$, since Eq. (\ref{abasym}) is
not valid for all $\theta$.
Eq. (\ref{abasym}) does however suggest a more appropriate way to define
the scattered wave.
We consider the definition:
\equation
	\psi_{\mbox{\scriptsize{scat}}} = \psi_{\mbox{\scriptsize{full}}}
	 -e^{-ikx -i\zeta\theta}
	\label{psiscatdef}
\endequation
where $\psi_{\mbox{\scriptsize{full}}}$ is the full wave function given in
Eq. (\ref{abwfn1}) which can be evaluated (for example) numerically.
This definition is not without problems since, of course,
$\psi_{\mbox{\scriptsize{scat}}}$
is not single--valued.
There are however several ways to circumvent this problem and to obtain useful
results with this approach.

The most aesthetically pleasing approach to dealing with this
``multi--valuedness'' problem is to work with the currents rather than
with the wave functions whenever possible. We begin by taking the covariant
derivative
of (\ref{abwfn1}) directly. This gives the simple result
\equation
	({\bf{\nabla}}-ie{\bf{A}})\psi = -{\bf{i}}ik\psi + {\bf{i}}f_x  +{\bf{j}}f_y,
	\label{covpsi}
\endequation
where
\begin{eqnarray}
	f_x & = & {k\over 2}[e^{i\zeta\pi/2}(e^{-i\theta}J_{-\zeta
+1}-iJ_{-\zeta})+e^{-i\zeta\pi/2}
	 (e^{-i\theta}J_{\zeta -1}+iJ_{\zeta})],\label{fx}\\
	f_y & = & {k\over 2}[e^{i\zeta\pi/2}(ie^{-i\theta}J_{-\zeta
+1}-J_{-\zeta})+e^{-i\zeta\pi/2}
		(ie^{-i\theta}J_{\zeta -1}+J_{\zeta})].\label{fy}
\end{eqnarray}
The Schr\"{o}dinger equation can now be used to derive an
equation satisfied by $f_x$ and $f_y$.  Applying the covariant derivative  to
(\ref{covpsi}) gives the relation
\equation
	-ikf_x+{\partial f_x \over\partial x}+{\partial f_y\over \partial y}
	  -ieA_x f_x -ieA_y f_y =0,
	\label{fcond}
\endequation
which may be verified explicitly.

For the case of a lattice of flux tubes the covariant derivative of the
total wave function may be obtained, to leading order in $\zeta$, by adding,
to the term $-{\bf i}ik\psi_{\rm total}$, the sum of the contributions of the
$f_x$'s and $f_y$'s from each
flux tube. We shall specialize to the case when there is only one column
of flux tubes and when $k_y$$=$$0$ (i.e. normal incidence).  We shall discuss
the more general
case later.  Let us define the vector ${\bf \Omega}(x,y)$
as the sum of the contributions  of $f_x$ and $f_y$ from all the
flux tubes in the lattice.  That is
\begin{eqnarray}
	\Omega _x (x,y) & = & \sum_{n=-\infty}^{\infty} f_x^n (kr_n ,\theta _n,\zeta
),
	\nonumber \\
	\Omega _y (x,y) & = & \sum_{n=-\infty}^{\infty} f_y^n (kr_n ,\theta _n,\zeta
),
	\label{omegadef}
\end{eqnarray}
where
\equation
	r_n = (x^2+(y-n\xi)^2)^{1/2},
\endequation
and where $\theta _n = \pi$ is parallel to the $x$--axis and corresponds to the
direction of propagation of the incident wave.  To lowest order in $\zeta$ the
total
wave function satisfies
\equation
	({\bf{\nabla}}-ie{\bf{A}})\Psi = (-{\bf{i}}ik) \Psi + {\bf{\Omega}}
	\label{covpsi2}
\endequation
This is analagous to what is done in ordinary scattering theory where the
scattered
waves from each individual atom are summed to obtain the total scattered wave.

To further check the self consistency of  the above procedure we
apply the covariant derivative to Eq. (\ref{covpsi2}). $\Psi$ is expected
to satisfy the Schr\"{o}dinger equation far away from any of the flux tubes.
(In fact in our case it will turn our that ``far away'' may be as close
as a few lattice spacings.)  It will do so
provided ${\bf \Omega}$ satisfies the equation
\equation
	-ik\Omega _x +{\partial\Omega_x \over \partial x} +{\partial\Omega_y\over
	\partial y} -ieA_x \Omega_x -ieA_y \Omega_y =0
	\label{omcond}
\endequation
where $A_x$ and $A_y$ are the total gauge potentials
given in (\ref{afieldx}) and (\ref{afieldy}).  Consider
the asymptotic region on the left where Eq. (\ref{ayasym}) holds.  Applying
(\ref{fcond})
to every term in the sums in (\ref{omcond}) gives
\equation
	i\zeta\pi\left [\sum_{n} \left ( {y\over \pi r_n^2}f_x^n -{x\over \pi r_n^2}
	   f_y^n \right ) -{\Omega _y\over \xi} \right ] = O(\zeta^2\pi^2).
	\label{omcond2}
\endequation
It is easy to see that the $f$'s are, asymptotically, of order $\zeta\pi$.
Thus the Schr\"{o}dinger equation will be satisfied asymptotically provided
$\bf\Omega$ is also of order $\zeta\pi$.
Eq. (\ref{omegadef}) guarantees,
that this will be the case.

We have not been able to obtain an analytic expression for $\bf\Omega$, even
asymptotically.
We have thus chosen to
evaluate ${\bf{\Omega}}$ numerically. Some details of the numerical work are
discussed
in the Appendix. We  have found that to a very high precision
and for a range of $k$ which is also discussed in the Appendix (basically
$2\pi > k\xi >> 2\zeta\pi$)
\equation
	{\bf{\Omega}}\simeq -{2i\over \xi}{\mbox{sin}}(\zeta\pi) ~e^{-ikx}{\bf j}.
	\label{omega}
\endequation
This result is valid provided one is not too close to the array of flux tubes.
Note also that to this order in $\zeta$ the value of $k$ (which is actually
$k_x$)
is the same for the incident and the transmitted wave.  Also, as expected,
when $\zeta\pi$ is small (\ref{omcond2}) is satisfied.

The next step in evaluating the scattering angle is to compute the current.
Using (\ref{covpsi2}) the current is found to be
\begin{eqnarray}
	{\bf{J}} & = & {1\over m}{\mbox{Im}}[\Psi^*( {\bf{\nabla}}-ie{\bf{A}})\Psi ],
\\
	         & = & {1\over m}[-{\bf{i}}k|\Psi|^2 +{\mbox{Im}}(\Psi^*
			{\bf{\Omega}})]
		\label{currentJ}
\end{eqnarray}
This expression could potentially cause us some difficulty since it involves
$\Psi$
directly whereas we have only computed the covariant derivative of $\Psi$.
Fortunately, to the order in $\zeta$ to which we are working, there is no
ambiguity in $\Psi$ and we can take $\Psi$ to be simply the incident wave
since linear order corrections will lead to $quadratic$ order corrections
to the scattering angle. Thus
\equation
	\Psi = e^{-ikx} + O(\zeta\pi)
	\label{incplcor}
\endequation
and
\equation
	|\Psi |^2 = 1 + O(\zeta\pi)
\endequation
We are now ready to use Eq. (\ref{currentJ}) to evaluate the scattering angle
$\phi_{\mbox{\scriptsize{trans}}}$.
\begin{eqnarray}
	{\mbox{tan}}\phi_{\mbox{\scriptsize{trans}}} & =& -{2\zeta\pi\over k\xi}
	+O(\zeta^2\pi^2)
\end{eqnarray}
Eqs. (\ref{deltaphi}) and (\ref{bmf})
can now be used to compare this result with the scattering off
of a strip of constant magnetic field.
We see that to lowest order in $\zeta\pi$ the scattering from an array
of flux tubes agrees exactly with the Mean Field result (which coincides
with the analytic result Eq. (\ref{phiexact}) in this case).  This
agreement provides a good check of our numerical method.

In the previous discussion we have avoided summing the non single--valued
scattered wave functions by summing only their covariant derivatives.
It is however possible to sum
the scattered waves directly
without first evaluating $D\psi$.
Using the definition (\ref{psiscatdef}) of the scattered wave
from an individual flux tube we note that the scattered wave at $(x,y)$ due to
a
flux tube located at $n\xi$ is given by
\equation
	\psi_{\mbox{\scriptsize{scat}}}^n (kr_n,\theta_n,\zeta)=
	(\psi (kr_n,\theta_n,\zeta )
	 -e^{-ikr_n{\mbox{\scriptsize{cos}}}\theta_n -i\zeta\theta_n})
	\label{psiscatn}
\endequation
where $\psi$ is the full wave function for each individual scatterer
and $r_n$ and $\theta_n$ are defined as in the previous discussion.
To lowest order in $\zeta$
the total scattered wave will  be the sum of the waves scattered
from each individual flux tube:
\equation
	\Psi _{\mbox{\scriptsize{scat}}} = \sum_{n=-\infty}^{\infty}
	\psi_{\mbox{\scriptsize{scat}}}^n (kr_n,\theta_n,\zeta).
	\label{psiscat}
\endequation
We know of no way to do this sum analytically.
We have thus evaluated (\ref{psiscat}) numerically by
summing the terms until satisfactory
convergence was achieved.
For $kr_n$ large and $\theta_n$ near $\pm \pi/2$ the asymptotic
form (\ref{abasym}) for  $\psi$ for each flux tube was used.
In other cases the sum (\ref{abwfn1}) was
evaluated directly.  This sum converges fairly rapidly for small and moderate
values of $kr_n$ since for large index
\equation
	J_n (x) \sim {1\over \sqrt{2\pi n}}({ex\over 2n})^n.
\endequation

The numerical work showed that to  the left of the scattering
region and for momenta in the range $2\pi$$>$$k\xi$$>>2\zeta\pi$
the scattered wave has the approximate form
\equation
	\Psi _{\mbox{\scriptsize{scat}}}
	\simeq e^{-ikx} (\epsilon(x,y) +i{\rm sin}(\zeta\pi)
		\chi (y)).
	\label{psiscat1}
\endequation
For $k\xi$ not too small $\epsilon$ was found to be negligible compared to
$\zeta\pi$
and $\chi$ was found to have the approximate form:
\equation
	\chi(y) \simeq -{2(y-n\xi)\over \xi} +{k\xi-1\over k\xi}
	~~~~~ {\rm for} ~~n\xi<y<(n+1)\xi
	\label{chi}
\endequation
for $\-\infty<n<+\infty$.
In fact $\chi$ does have some
oscillatory $x$ dependence but it was  found to be negligible
compared to the  $y$ dependence of $\chi$.
The slope  ``$-2/\xi$'' is accurate to
about $.2\%$, and  $(k\xi-1)/k\xi$ is accurate to about
$1\%$.  For a more thorough discussion of $\epsilon$ and $\chi$ and of the
limits on the momentum we again refer the reader to the Appendix.

It is clear from our method that $\Psi _{\mbox{\scriptsize{scat}}}$ must have
discontinuities along lines
starting at the site of each flux tube  and extending out to the left parallel
to the $x$ axis.  This is because we have subtracted out the multi--valued
incident wave exp($-ikx-i\zeta
\theta$) from each (single--valued) wave function.
This brings up the difficult problem of what to choose as
the $incident$ wave in the multi--vortex case. Although it is true that
to lowest order it is simply a plane wave we must be very careful
since the scattered wave (as we have defined it) is not single valued.

One important condition which $\Psi_{\mbox{\scriptsize{inc}}}$ should
satisfy is
\equation
	({\bf{\nabla}}-ie{\bf{A}})\Psi_{\mbox{\scriptsize{inc}}} = -{\bf i}ik
		\Psi_{\mbox{\scriptsize{inc}}}
	\label{covder}
\endequation
The form $\Psi_{\mbox{\scriptsize{inc}}}=$exp$(-ikx+i\zeta\pi y/\xi )$
satisfies this condition asymptotically (when $x$$<<$$0$).
However due to the form of $\chi$ which
has discontinuities resulting from its multi--valuedness
it is important to choose $\Psi_{\mbox{\scriptsize{inc}}}$ to have
the same discontinuities. In fact
\equation
	\Psi_{\mbox{\scriptsize{inc}}}=e^{-ikx}\left(1 + i\zeta\pi{(y-n\xi)
		\over\xi}\right) ~~~~~ {\rm for} ~~n\xi<y<(n+1)\xi
\endequation
also satisfies Eq. (\ref{covder}) asymptotically but only to order
$\zeta\pi$.
Setting
\equation
	\Psi = \Psi_{\mbox{\scriptsize{inc}}}+\Psi _{\mbox{\scriptsize{scat}}},
	\label{psii+s}
\endequation
we can compute the covariant derivative of the $total$ wave function $\Psi$.
To linear order in $\zeta\pi$
\equation
	({\bf{\nabla}}-ie{\bf{A}})\Psi = -{\bf i}ik\Psi -{\bf j}
	     {2i\over \xi}\zeta\pi e^{-ikx}
\endequation
This is precisely the result which we obtained using our previous method
(see Eqs. (\ref{covpsi2}) and (\ref{omega}) ).
We thus see that with this method as well, we recover the
Mean Field result when the flux per particle is small.
Note that we have {\bf shown} that $\Psi$ is a plane wave with corrections
linear in $\zeta\pi$. This was  {\bf assumed} in our
previous discussion (Eq. (\ref{incplcor})).

It likely that the linear
correction to the plane wave above should be
exponentiated.  From this point of view the discontinuities
are very useful since they allow the perturbative corrections
to $\psi$ to remain small even for large $y$.

It is by now clear that the two approaches discussed above are
complementary.  In the former approach it appeared as though we
could avoid any problems of discontinuities but in fact the
discontinuities were all ``hidden'' in $\Psi$.  The latter approach
has the advantage of showing clearly what is going on by evaluating
the wave function directly. From a calculational point of view the
former approach is advantageous (at least in lowest order) since
the derivative of $\Psi$ is found directly without having to
evaluate $\Psi$ at several points and perform a linear regression
to find the slope.

It is a simple matter to extend the above results to the case where
$k_y$ is nonzero.  A subtlety which one encounters numerically
is that the wave functions (or their covariant derivatives)
due to each flux tube
must be multiplied by appropriate phases in order that the plane waves
incident at each site are in phase. Our two methods again give complementary
results so we describe here only the ``first'' method in which the
covariant derivative or $\Psi$ is taken first.  The appropriate generalization
of (\ref{covpsi2}) is
\equation
	({\bf \nabla} -ie{\bf A})\Psi = (-{\bf i}ik_x+{\bf j}
		ik_y)\Psi +{\bf\Omega},
\endequation
where ${\bf \Omega}$ has been studied  numerically and
is approximately given by
\equation
	{\bf\Omega}\simeq-{2i\over\xi}sin(\zeta\pi)e^{i(k_yy-k_xx)}\left(
		{2k_y\over k_x}{\bf i}+{\bf j}\right)
\endequation
in the asymptotic region on the left.  We refer the reader to
the Appendix for more details of the numerical results.
Generalizing from the result for $k_y=0$ we take
\equation
	\Psi = e^{i(k_yy-k_xx)}+O(\zeta\pi)
\endequation
We then find
\equation
	{\mbox{tan}}(\phi_{\mbox{\scriptsize{trans}}}-\phi_{\mbox{\scriptsize{inc}}})
		=-{2\zeta\pi\over k_x \xi} +O(\zeta^2\pi^2).
\endequation
This result again agrees with the Mean Field result (Eq. (\ref{mainmf})) and
with
the exact result, since ``$m$''$=$$0$.

The extension of our results to the case where there are several columns is
also a simple matter provided the perturbative analysis remains valid.
Suppose we were to add a new column positioned at
$x$$=$$\xi$.  In the local coordinates ($x^{'},y^{'}$) of that column an
incident plane
wave exp($ik_yy^{'}-ik_xx^{'}$) gives rise to
\equation
	{\bf\Omega ^{'}}=-{2i\over\xi}sin(\zeta\pi)e^{i(k_yy^{'}-k_xx^{'})}\left(
		{2k_y\over k_x}{\bf i}+{\bf j}\right).
\endequation
Since $y^{'}$$=$$y$ and $x^{'}$$=$$x-\xi$ the wave incident on the
column at $x$$=$$\xi$ must be multiplied by the phase exp($-ik_x\xi$)
to obtain the correct
phase of the  plane wave at the locations of the two columns.
Thus
\equation
	{\bf\Omega}_{\mbox{\scriptsize{tot}}}={\bf\Omega}+e^{-ik_x\xi}
		{\bf\Omega^{'}} = 2{\bf\Omega}
\endequation
so that
\equation
	{\mbox{tan}}(\phi_{\mbox{\scriptsize{trans}}}-
		\phi_{\mbox{\scriptsize{inc}}})
		=-{4\zeta\pi\over k_x \xi} +O(\zeta^2\pi^2)
\endequation
for $N$$=$$1$.  In general, for $N+1$ columns,
\equation
	{\mbox{tan}}(\phi_{\mbox{\scriptsize{trans}}}-
		\phi_{\mbox{\scriptsize{inc}}})
		=-{2(N+1)\zeta\pi\over k_x \xi} +O(\zeta^2\pi^2).
\endequation
The net effect is that our perturbative expansion is now in ``$(N+1)\zeta\pi$''
rather than in ``$\zeta\pi$''. Thus even for multiple columns
the scattering angle agrees with the exact result and thus with
that obtained from the Mean Field
analysis provided the flux per particle is sufficiently small.  Again
this agreement provides a good check of our numerical method.

So far we have concentrated on comparing the {\bf scattering angle}
for the lattice of flux tubes with that obtained in the Mean Field analysis.
We  now describe a comparison of another quantity of interest namely
the transmission and reflection amplitudes $\tau$ and $\rho$, for which
there are no exact results available.
We shall discuss only the case  $k_y$$=$$0$ for which
we shall compute numerically the sum in Eq. (\ref{psiscat}).

In the previous section we computed $\rho$ and $\tau$ for the
Mean Field case. In terms of the ``flux parameters'' $\zeta$
and $\xi$ and in the limit
when $\zeta\pi$ is small and $k\xi$$>>$$|\zeta |\pi$ they are given by (see Eqs
(\ref{bmf}), (\ref{dmf}),
(\ref{rho}) and (\ref{tau})):
\begin{eqnarray}
	\rho_{\mbox{\scriptsize{mf}}} & = & {\rm e}^{ik\xi}
	    \zeta ^2 \pi ^2\left( {{\mbox{sin}}^2k\xi\over (k\xi )^4}
		+i ({{\mbox{sin}}2k\xi\over 2(k\xi )^4 } -{1\over
		 (k\xi )^3})\right)
		+ O(\zeta ^3\pi ^3),\label{rhomf}\\
	\tau_{\mbox{\scriptsize{mf}}} & = & 1 + \zeta ^2\pi ^2\left(
		{1\over (k\xi )^2}-{2\over 3}{i
		\over k\xi }\right) +O(\zeta ^3\pi ^3). \label{taumf}
\end{eqnarray}
(The extra factor of ${\rm e}^{ik\xi}$ has been added to compensate
for the  positioning of the mean field relative to the
flux tubes.)
Note that to order $\zeta\pi$ we have the simple result that
$\rho$$=$$0$ and $\tau$$=$$1$. To this order the wave is entirely
transmitted in agreement with the {\bf classical} result.
We now describe the calculation for $\rho$ and $\tau$ for the lattice
of flux tubes.

We begin by summing (\ref{psiscat}) numerically. When $k\xi$$>>$$2|\zeta |\pi$
and $x$$>$$0$ we find
\equation
	\Psi _{\mbox{\scriptsize{refl}}} = -{i\over k\xi} {\rm sin}\zeta\pi
		 e^{ikx} \label{psirefl},
\endequation
to an accuracy of about $1 \%$ in the coefficient of ``$-i/k\xi$''.
A naive interpretation of this result would imply
\equation
	\rho_{\mbox{\scriptsize{ft}}} = -{i\zeta\pi\over k\xi} +O(\zeta^2\pi^2),
	\label{rhowrong}
\endequation
which is {\bf linear} in $\zeta\pi$. This, of course, {\bf differs}
from the Mean Field result.
We now argue that this naive interpretation is incorrect and that
to order $\zeta\pi$ we must be more careful in defining the
reflection and transmission amplitudes. Physically this is due
to the fact that the $angle$ of the wave has been changed to order
$\zeta\pi$ by
the scattering. This change in angle should $not$ affect the
transmission amplitude. More concretely recall that
to the left of the scattering region
the scattered
wave has the form
\equation
	\Psi _{\mbox{\scriptsize{scat}}} =  i{\rm sin}(\zeta \pi)
		 \chi(y) e^{-ikx}
	\label{psiscatsummed}
\endequation
($\epsilon$ is negligible for large $k\xi$)  and that $\chi (y)$ is a
``sawtooth'' with discontinuities whenever $y$$=$$n\xi$.
We have  discussed previously that $\chi$ must somehow exponentiate
to give the correct plane wave in the $y$ direction.
It is thus incorrect to read off the transmission coefficient
naively from Eq. (\ref{psiscatsummed}) which would give
\equation
	\tau_{\mbox{\scriptsize{ft}}}=
	1 + i\zeta\pi \bar{\chi}+O(\zeta^2\pi^2),
	\label{tauwrong}
\endequation
In fact the contribution of $\zeta$ is simply to modify the
plane wave to take into account the ``bending'' of the particle
in the magnetic field and the $correct$ $\tau$ to this order is {\bf one}.
Thus the apparent $O(\zeta\pi)$ correction
to the transmission amplitude is
best understood as  an artifact of our approach and the difficulties which
the Aharonov--Bohm phases cause
when one tries to add up the scattered waves.

Once we realize that $\tau$ is one to order $\zeta\pi$ it is
clear that $\rho$ must vanish to this order. This follows
from flux conservation.
The fact that the  $\rho_{\mbox{\scriptsize{ft}}}$ which we {\bf naively}
calculated is of order $\zeta\pi$ is a direct consequence of the
fact that the ``naive'' $\tau_{\mbox{\scriptsize{ft}}}$
has a piece which  is linear in $\zeta\pi$ since
\equation
	|\rho_{\mbox{\scriptsize{ft}}}
	 |^2 +{{\tilde k}\over k}|\tau_{\mbox{\scriptsize{ft}}}|^2 ~~\simeq ~~
		{\zeta ^2 \pi ^2\over k^2\xi^2} +(1 - {2\zeta ^2\pi ^2\over
		k^2\xi^2})(1+{\zeta ^2\pi ^2\over k^2\xi^2})
	 	~~ \simeq ~~ 1
\endequation
where we have used the fact (from Eq. (\ref{chi})) that
\equation
	\bar{\chi (y)} \equiv \chi (y=.5\xi ) \simeq -{1 \over k\xi},
\endequation
It may of course happen that
$\tau_{\mbox{\scriptsize{ft}}}$ has real pieces quadratic in $\zeta\pi$
(to which our calculation would  be insensitive). In this case these
should be included
in $|\tau_{\mbox{\scriptsize{ft}}} |^2$. In fact the above argument
used numerical results and it is thus just valid in
an approximate sense. In any case it is evident
that the apparent discrepancy between the form of $\rho$ and $\tau$
in the ``flux tube'' and the Mean Field case is simply an
 artifact of our approach.

\section{Discussion}

In this paper we set out to understand the validity of Mean Field
Theory when applied to scattering from infinitesimally small flux
tubes for which there is no Lorentz force except possibly at the
location of the flux tubes. We considered the scattering of a plane
wave representing an incident charged particle from an array of flux
tubes with infinite extent in the $y$ direction. By computing
the current of the wave beyond the scattering region
we found that the incident beam {\bf was} bent by the array of flux
tubes. When the flux per tube is small the angle by which the
beam is bent is given precisely by the Mean Field result.

Caenepeel and MacKenzie \cite{cm1} have recently shown that no bending
occurs when scattering from a single flux tube unless an additional
interaction besides the
Aharonov--Bohm interaction is present \cite{mrw}. From our
calculation one can see that even in the absense of an additional
interaction there is precisely the correct (MF) amount of bending
from an {\bf array} of flux tubes. Although our calculation was
done (for essential technical reasons) for a regular lattice,
it seems very likely on physical grounds that the main results will persist for
a
random array of flux tubes.

Despite the absense of a Lorentz force for a lattice of flux tubes
there is a way to understand how such bending  can occur $classically$.
Imagine first taking each flux tube to be of finite but small
size, computing the scattering, and then letting the size become very small
keeping the mean field, and thus the flux per tube, constant.
One finds that the charged particle will have to travel an increasingly
large distance before scattering but, after many interactions,
 the mean impulse it receives from
all the tubes from which it scatters is precisely that which it would
have received from the {\bf mean} field. There is however a catch.
This only works when the flux tubes are larger than some critical
size at which the magnetic field inside the flux tube becomes
so large that the radius of the path of the particle is equal to the
radius of the flux tube.  Beyond this critical size, the
average deflection
which the particle receives becomes too small.
It is only  Quantum Mechanical scattering to which  the
Mean Field result truly applies when the flux per tube is small.

\section{Appendix: Comments on the Numerical Work}

In this appendix we describe several interesting features of
our numerical work which were not mentioned in the text.

We begin by discussing the numerical work leading to
Eq. (\ref{psiscat1}) for $\Psi_{\mbox{\scriptsize{scat}}}$.
Recall that for $k_y$$=$$0$ we claimed that our numerical results
lead to a scattered wave of the form
\equation
	\Psi_{\mbox{\scriptsize{scat}}} \simeq  e^{-ikx} (\epsilon +
	i{\mbox{sin}}(\zeta\pi)\chi (y)),
	\label{ap1}
\endequation
with
\equation
	\chi(y) \simeq -{2(y-n\xi)\over \xi} +{k\xi-1\over k\xi}
	\label{ap2}
\endequation
in the region to the left of the flux tubes.
The result (\ref{ap1}) was obtained by summing the series
in Eq. (\ref{psiscat}) numerically
until satisfactory convergence was achieved.  For very small $k\xi$ this
involved adding the contributions from up to a million flux tubes
in order to obtain a slope in $\chi$ accurate to about .2\%.
As one might expect from the
fact that the vector potential approaches the MF result exponentially (c.f.
Eqs. (\ref{afieldx}) - (\ref{ayasym}) ) the above formula becomes valid
within a very short distance of the scattering region (i.e. within
a distance of about $4*\xi$).

For $k\xi$ very large compared to $\zeta\pi$, $\epsilon$ was negligible.
For $k\xi$ closer to $\zeta\pi$ the maximum value of $|\epsilon |$ was
found to be approximately proportional to $\zeta\pi$ and to have an
approximate $1/k\xi$ dependence.  From (\ref{ap2}) $\chi$ is also
dominated by a $1/k\xi$ dependence for small $k\xi$.  For all cases
investigated (we examined $\zeta$ as low as $.0001$ and $k\xi$ as
low as $.005$) it has been found that
\equation
	|\epsilon |<<{\rm sin} (\zeta\pi )|\chi |,
\endequation
for small $k\xi$.

The growth of $\chi$ with small $k\xi$ has an interesting interpretation.
{}From a MF point of view we would expect the particle to be
totally reflected for $k\xi$$<$$2|\zeta |\pi$.  From our perturbative
approach we see that the results become non--perturbative when
$\zeta\pi\chi$ becomes of order $1$ i.e. when $k\xi$$\sim$$|\zeta |\pi$.
In a loose sense our approach predicts the correct ``critical'' point.

Numerical work was also required for evaluating $\bf\Omega$ defined
in Eq. (\ref{omegadef}).
In order to determine $\bf\Omega$ for the case  $k_y$$=$$0$ the exact
forms of $f_x$ and $f_y$
were used for small $kr$.  For larger $kr$ approximate forms (easily
derived from (\ref{fx}) and (\ref{fy})) were employed.  As in the previous
calculation it was necessary to sum  up to a
million tubes.  For $k\xi$ very large compared to $2\zeta\pi$
and slightly less than $2\pi$ the form quoted for $\Omega_y$ in
(\ref{omega}) was accurate to about $.2\%$.  Similarly,  $\Omega_x /\Omega_y$
was ``zero'' to a magnitude of about $.002$.

For the more general case where $k_y$ is nonzero, things are much the
same.  A small  difference was that some precision was lost in
$\Omega_x$ when $k_y$$<<$$k_x$.  The reason is that the number
which was expected was proportional to $k_y/k_x$ and thus became very small.

For large momenta it was found that
$\Psi_{\mbox{\scriptsize{scat}}}$ and ${\bf \Omega}$ deviated from the forms
quoted in the main body of the paper due to diffraction effects.  Recall that
diffraction occurs when circular waves ``emanating'' from neighbouring sites
add up coherently.  For small momenta there is only one angle at which
this occurs and only the ``forward peak'' is observed.  For higher
momenta it becomes possible to have coherence in more than one direction.
In effect, the transmitted wave becomes the sum of more than one
plane wave.  As discussed near the end of Section III.A, the scattering
angle then becomes ill defined.

One can derive the limit on
$k_x$ and $k_y$ by using Eq.(\ref{zetapm}) for the ``critical'' fluxes.
Since our numerical approach is only valid near $\zeta$$=$$0$, we shall
for convenience set $\zeta$$=$$0$ for this derivation.  Diffraction
then starts when either $\zeta^{+}_{-1}$ or $\zeta^{-}_{+1}$ is equal
to zero so that the limit on the momenta is easily seen to be
\equation
	(k_x^2+k_y^2)^{1\over 2}\xi+|k_y|\xi < 2\pi .
	\label{diffrac}
\endequation

\vskip .3in
\centerline {\bf Acknowledgements}
\bigskip
This work is supported in part by the Natural Sciences and Engineering
Research Council of Canada. Their support is gratefully acknowledged.
We wish to thank Ian Kogan, Arnold Sikkema and Stephanie Curnoe for
helpful discussions.

\end{document}